\documentclass[onecolumn,showpacs]{revtex4}

\usepackage{graphicx}% Include figure files
\usepackage{dcolumn}% Align table columns on decimal point
\usepackage{bm}% bold math

\input epsf

\begin{document}

\title{\Large Acceleration of the Universe in Presence of Tachyonic field}

\author{\bf  Surajit
Chattopadhyay$^1$\footnote{surajit$_{_{-}}2008$@yahoo.co.in},
Ujjal Debnath$^2$\footnote{ujjaldebnath@yahoo.com ,
ujjal@iucaa.ernet.in} and Goutami Chattopadhyay$^2$}

\affiliation{$^1$Department of Information Technology, Pailan
College of Management and Technology, Bengal Pailan Park,
Kolkata-700 104, India.\\
$^2$Department of Mathematics, Bengal Engineering and Science
University, Shibpur, Howrah-711 103, India. }

\date{\today}

\begin{abstract}
In this letter, we have assumed that the Universe is filled in
tachyonic field with potential, which gives the acceleration of
the Universe. For certain choice of potential, we have found the
exact solutions of the field equations. We have shown the
decaying nature of potential. From recently developed statefinder
parameters, we have investigated the role of tachyonic field in
different stages of the evolution of the Universe.
\end{abstract}


\maketitle

Recent observations on microwave background radiation have
offered good amount of support to cosmological inflation to be an
integral part of the standard model of the universe. On the other
hand there have been difficulties in obtaining accelerated
expansion from fundamental theories such as M/String theory [1].
Much has been written and emphasized about the role of the
fundamental dilation field in the context of string cosmology.
But, not much emphasized is tachyon component [2]. It has been
recently shown by Sen [3, 4] that the decay of an unstable
D-brane produces pressure-less gas with finite energy density
that resembles classical dust. The cosmological effects of the
tachyon rolling down to its ground state have been discussed by
Gibbons [5]. Rolling tachyon matter associated with unstable
D-branes has an interesting equation of state which smoothly
interpolates between $-1$ and 0. As the Tachyon field rolls down
the hill, the universe experiences accelerated expansion and at a
particular epoch the scale factor passes through the point of
inflection marking the end of inflation [1]. The tachyonic matter
might provide an explanation for inflation at the early epochs
and could contribute to some new form of cosmological dark matter
at late times [6]. Inflation under tachyonic field has also been
discussed in ref. [2, 7, 8]. Sami et al [9] have discussed the
cosmological prospects of rolling tachyon with exponential
potential. In this letter, we have assumed that the Universe is
filled in tachyonic field with potential, which
gives the acceleration of the Universe.\\

The action for the homogeneous tachyon condensate of string
theory in a gravitational background is given by,
\begin{equation}
S=\int {\sqrt{-g}~ d ^{4} x \left[\frac{\cal R}{16 \pi G}+{\cal
L}\right]}
\end{equation}
where $\cal L$ is the Lagrangian density given by,
\begin{equation}
{\cal {L}}=-V(\phi)\sqrt{1+g^{\mu \nu}~\partial{_{\mu}}\phi
\partial{_{\nu}} \phi}
\end{equation}
where $\phi$ is the tachyonic field, $V(\phi)$ is the tachyonic
field potential and $\cal R$ is the Ricci Scalar. The
energy-momentum tensor for the tachyonic field is,

\begin{eqnarray}
\begin{array} {ccc}
T_{\mu \nu}=-\frac{2 \delta S}{\sqrt{-g}~ \delta g^{\mu
\nu}}=-V(\phi)\sqrt{1+g^{\mu \nu} \partial _{\mu}\phi
\partial_{\nu} \phi}g^{\mu \nu}+V(\phi) \frac{\partial _{\mu}\phi
\partial_{\nu}
\phi}{\sqrt{1+g^{\mu \nu} \partial _{\mu}\phi \partial_{\nu} \phi}}\\\\\
=p~g_{\mu \nu}+(p+\rho)u_{\mu} u_{\nu}\\\\
\end{array}
\end{eqnarray}

where the velocity $u_{\mu}$ is :
\begin{equation}
u_{\mu}=-\frac{\partial_{\mu}\phi}{\sqrt{-g^{\mu \nu} \partial
_{\mu}\phi
\partial_{\nu} \phi}}
\end{equation}

with $u^{\nu} u_{\nu}=-1$.\\

The energy density $\rho$ and the pressure $p$ of the tachyonic
field therefore are,
\begin{equation}
\rho=\frac{V(\phi)}{\sqrt{1-{\dot{\phi}}^{2}}}
\end{equation}
and
\begin{equation}
p=-V(\phi) \sqrt{1-{\dot{\phi}}^{2}}
\end{equation}

Now the metric of a spatially flat isotropic and homogeneous
Universe in FRW model is

\begin{equation}
ds^{2}=dt^{2}-a^{2}(t)\left[dr^{2}+r^{2}(d\theta^{2}+sin^{2}\theta
d\phi^{2})\right]
\end{equation}

where $a(t)$ is the scale factor.\\

Here we have assumed that the universe is filled in only tachyonic
field, so the  Einstein field equations are (choosing $8\pi
G=c=1$) given by

\begin{equation}
3\frac{\dot{a}^{2}}{a^{2}}=\rho
\end{equation}
and
\begin{equation}
6\frac{\ddot{a}}{a}=-(\rho+3p)
\end{equation}

The energy conservation equation is
\begin{equation}
\dot{\rho}+3\frac{\dot{a}}{a}(\rho+p)=0
\end{equation}

which leads to

\begin{equation}
\frac{\dot{V}}{V\dot{\phi}^{2}}+\frac{\ddot{\phi}}{\dot{\phi}}
\left(1-\dot{\phi}^{2}\right)^{-1}+3\frac{\dot{a}}{a}=0
\end{equation}

Now, in order to solve the equation (11), we take a simple form of
$V=\left(1-\dot{\phi}^{2}\right)^{-m}, ~(m>0)$, so that the
solution of $\phi$ becomes

\begin{equation}
\phi=\frac{2a^{3/2}}{\sqrt{3c}}~~_{2}F_{1}
[\frac{1+2m}{4},\frac{3+2m}{4},\frac{5+2m}{4},-c^{-\frac{2}{2m+1}}a^{\frac{6}{1+2m}}]
\end{equation}
where $c$ is an integration constant. From figure 1, we see that
$\phi$ increases with $a$. The potential $V$ of the tachyonic
field $\phi$ can be written as

\begin{equation}
V=\left[1+\left(\frac{c}{a^{3}} \right)^{\frac{2}{1+2m}}
\right]^{m}
\end{equation}

So from equations (12) and (13), we have the relation between
$\phi$ and $V$ as

\begin{equation}
\phi=\frac{1}{\sqrt{3}}\left(V^{\frac{1}{m}}-1\right)^{\frac{1}{1+1m}}~~_{2}F_{1}
[\frac{1+2m}{4},\frac{3+2m}{4},\frac{5+2m}{4},-(V^{\frac{1}{m}}-1)^{-1}]
\end{equation}

From above expression of $V$, we see that $V$ decreases from large value to 1 as $a$ increases,
which is shown in graphically in figure 2. From figure 3, we see that $V(\phi)$ decreases as $\phi$ increases i.e., $V$ decreases
with evolution of the universe.\\

Also from equation (8), we have the solution for $a$ as

\begin{equation}
t=\frac{2}{\sqrt{3c}}~a^{\frac{3}{2}}~~_{2}F_{1}[\frac{1+2m}{4},\frac{1+2m}{4},\frac{5+2m}{4},-c^{-\frac{2}{2m+1}}a^{\frac{6}{1+2m}}]
\end{equation}

\begin{figure}
\includegraphics[height=1.9in]{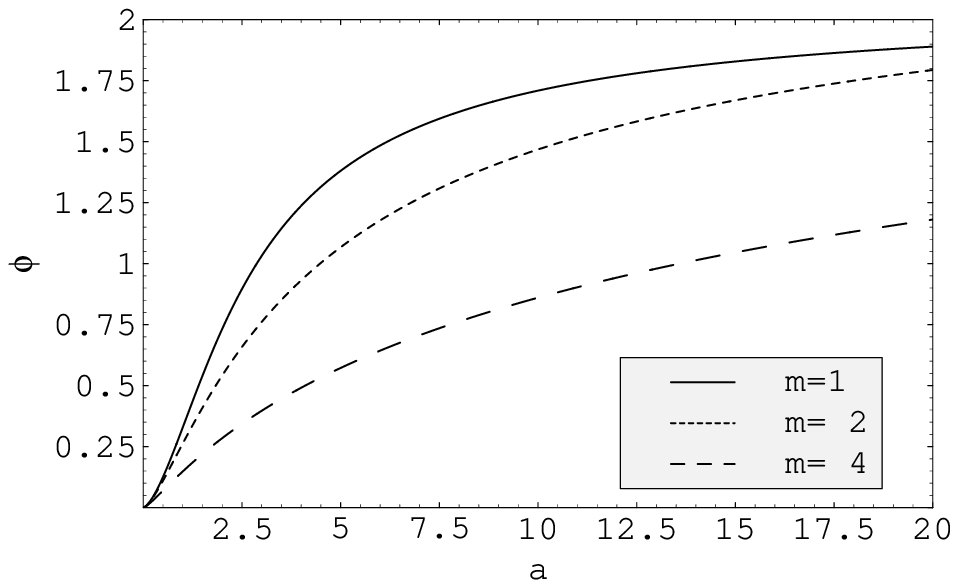}~~~
\includegraphics[height=1.9in]{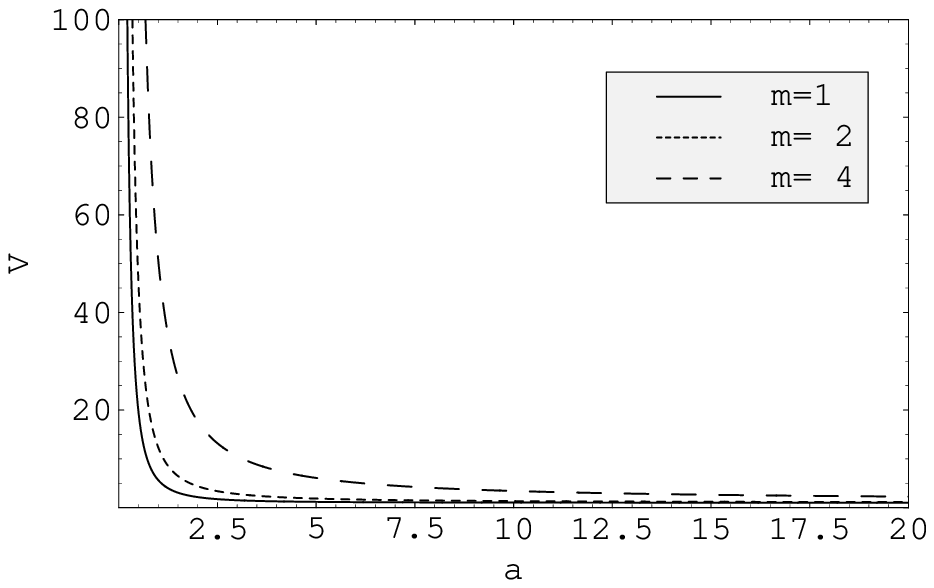}\\
\vspace{1mm}
Fig.1~~~~~~~~~~~~~~~~~~~~~~~~~~~~~~~~~~~~~~~~~~~~~~~~~~~~~~~~~~~Fig.2\\
\vspace{5mm}

\includegraphics[height=1.9in]{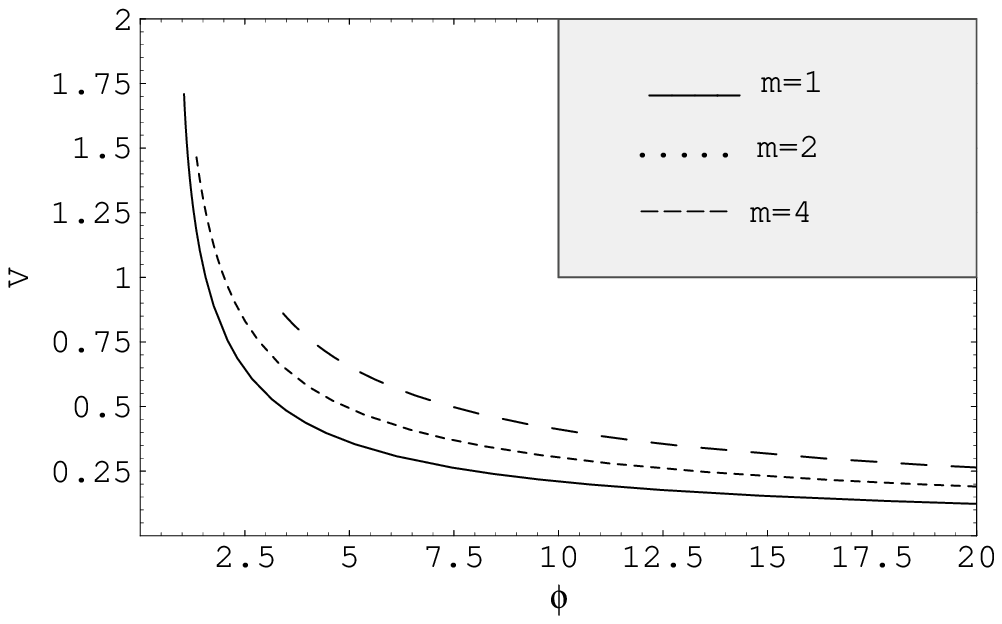}~~~
\includegraphics[height=1.9in]{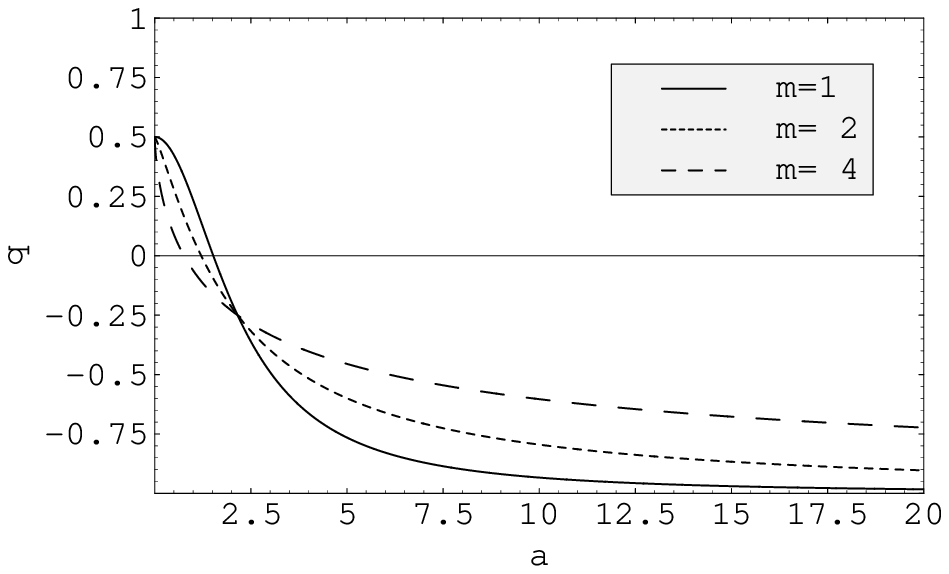}\\
\vspace{1mm}
Fig.3~~~~~~~~~~~~~~~~~~~~~~~~~~~~~~~~~~~~~~~~~~~~~~~~~~~~~~~~~~~Fig.4\\
\vspace{5mm} Figs. 1, 2 and 4 show variation of $\phi,~V$ and $q$
against $a$ respectively and fig. 3 shows variation of $V$ against
$\phi$ for different values of $m~(=~1,~2,~4)$. \hspace{2cm}
\vspace{6mm}

\end{figure}

The deceleration parameter $q$ has in the form

\begin{equation}
q=-\frac{a\ddot{a}}{\dot{a}^{2}}=\frac{1}{2}\left[1-\frac{3}{1+\left(\frac{c}{a^{3}}
\right)^{\frac{2}{1+2m}}} \right]
\end{equation}

For accelerating universe, $q$ must be negative i.e.,
$a>c^{\frac{1}{3}}~2^{-\frac{2m+1}{6}}$. From figure 4, we see
that $q$ decreases from 0.5 to $-1$ i.e., the sign flip from
positive to negative signature of $q$ in matter dominated era which is caused by tachyonic field.\\

From equations (5) and (6), we have

\begin{equation}
w_{eff}=\frac{p}{\rho}=-1+\dot{\phi}^{2}
\end{equation}

For real values of $\rho$ and $p$, we must have $\dot{\phi}^{2}\le
1$, which implies $-1\le w_{eff}\le 0$. This interprets that the
tachyonic field interpolates between dust and $\Lambda$CDM stages.
From figure 1, we see that the deceleration parameter $q$ lies
between 0.5 and $-1$ for different values of $m$ i.e., universe
starts from dust to $\Lambda$CDM model.\\

The statefinder diagnostic pair $\{r,s\}$, introduced by Sahni et
al [10] is constructed from the scale factor $a(t)$ and its
derivatives up to the third order as follows:

\begin{equation}
r=\frac{\dddot{a}}{aH^{3}}~~~~\text{and}~~~~s=\frac{r-1}{3\left(q-\frac{1}{2}\right)}
\end{equation}

where $H\left(=\frac{\ddot{a}}{a}\right)$ is the Hubble
parameter. These parameters are dimensionless and allow us to
characterize the properties of dark energy in a model independent
manner. The parameter $r$ forms the next step in the hierarchy of
geometrical cosmological parameters after $H$ and $q$. From
equations (16) and (18) we have the relation between $r$ and $s$:

\begin{equation}
8(2m-1)(r-1)^{2}+6(14m-5)(r-1)s-9(1+2m)(2r-5)s^{2}=0
\end{equation}

\begin{figure}
\includegraphics[height=2.7in]{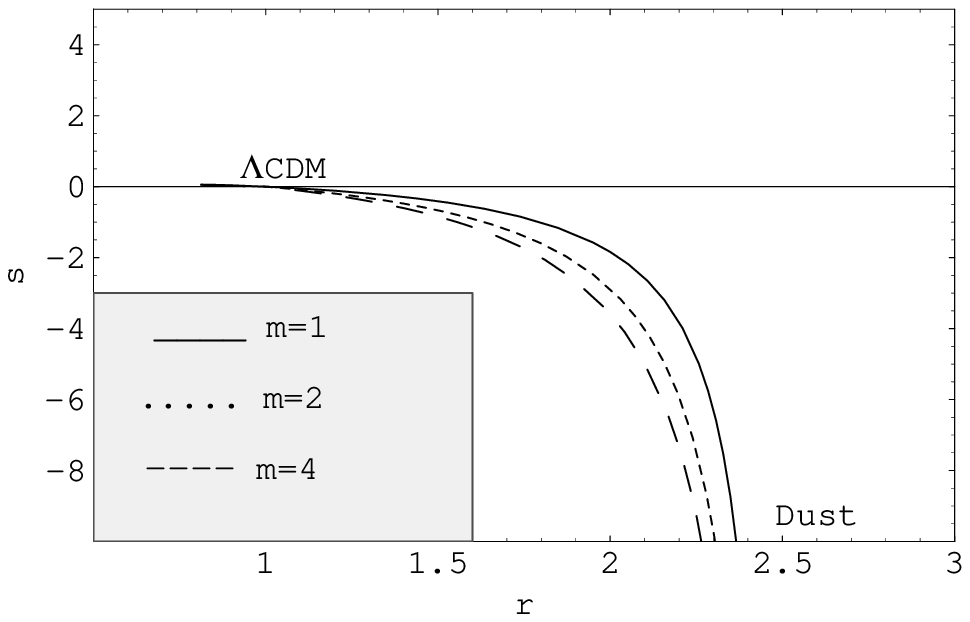}\\
\vspace{1mm} Fig.5\\

\vspace{5mm} Fig. 5 shows the variation of $s$
 against $r$ for different values of $m~(=~1,~2,~4)$.\hspace{2cm} \vspace{4mm}

\end{figure}

Figure 5 represents the variation of $s$ against $r$ for
different values of $m~(=~1,~2,~4)$. The negative side of $s$
represents the evolution of the universe starts from dust state
($r=$ finite, $s\rightarrow -\infty$) to the $\Lambda$CDM  ($r=1,~s=0$) model.\\

In this letter, we have considered the flat FRW Universe driven by
only tachyonic field. We have presented accelerating expansion of
our Universe due to tachyonic field. We found exact solutions of
tachyonic field and potential by considering specific form of
potential. We show that the potential represent the decaying
nature as $\phi$ or $a$ increases. Note that if $V=$ constant,
the tachyonic field represents the pure Chaplygin gas. If $V\ne$
constant, the tachyonic field may be treated as variable
Chaplygin gas. Graphical representation of $q$ shows that the
tachyonic field propagates between dust (early stage) and
$\Lambda$CDM stages (late stage). $\{r,s\}$ figure shows the
evolution of the universe starts from dust and ends at
$\Lambda$CDM stage. At whole stages of the evolution, $r$ is
always $\ge 1$ and $s$ becomes negative. Thus from the behaviour
of $q$ and $\{r,s\}$, we say that acceleration is possible if the
universe is filled in only tachyonic field.\\\\\\\\\

{\bf Acknowledgement:}\\\\
One of the authors (UD) is thankful to BESU, India for
providing a research project grant (No. DRO-2/6858).\\\\

{\bf References:}\\
\\
$[1]$ M. Sami, {\it Mod. Phys. Lett. A} {\bf 18} 691 (2003); [hep-th/0205146].\\
$[2]$ A. Feinstein, {\it Phys. Rev. D} {\bf 66} 063511 (2002); [hep-th/0204140].\\
$[3]$ A. Sen, {\it JHEP} {\bf 0204} 048 (2002); [hep-th/0203211].\\
$[4]$ A. Sen, {\it JHEP} {\bf 0207} 065 (2002); [hep-th/0203265].\\
$[5]$ G. W. Gibbons, {\it Phys. Lett. B} {\bf 537} 1 (2002); [hep-th/0204008].\\
$[6]$ M. Sami, P. Chingangbam and T. Qureshi, {\it Phys. Rev. D} {\bf 66} 043530 (2002); [hep-th/0205179].\\
$[7]$ M. Fairbairn and M.H.G. Tytgat, {\it Phys. Lett. B} {\bf 546} 1 (2002); [hep-th/0204070].\\
$[8]$ T. Padmanabhan, {\it Phys. Rev. D} {\bf 66} 021301 (2002); [hep-th/0204150].\\
$[9]$ M. Sami, P. Chingangbam and T. Qureshi, {\it Pramana} {\bf 62} 765 (2004); [hep-th/0301140].\\
$[10]$ V. Sahni, T. D. Saini, A. A. Starobinsky and U. Alam, {\it
JETP Lett.} {\bf 77} 201 (2003).\\

\end{document}